\documentclass[aps,pra,epsfigure,twocolumn,longbibliography]{revtex4-1}
\usepackage{dcolumn}    
\usepackage{bm} 
\usepackage{float}
\usepackage{graphicx}
\usepackage{caption}
\usepackage{amsmath}    
\usepackage{latexsym}
\usepackage{amsfonts}   
\usepackage{amssymb}
\usepackage{array}      
\usepackage{epsfig}
\usepackage{txfonts}
\usepackage{color}
\usepackage[dvipsnames]{xcolor}
\usepackage[colorlinks=true,linkcolor=blue,urlcolor=blue,citecolor=blue,pdfusetitle]{hyperref}
\usepackage{subcaption}
\usepackage{hyperref}
\usepackage{cleveref}

\usepackage{tikz}

\usetikzlibrary{calc,arrows}
\newcommand{\tikzmark}[1]{%
  \tikz[overlay,remember picture] \coordinate (#1) {};}

\usetikzlibrary{arrows,positioning} 
\tikzset{
    >=stealth',
    punkt/.style={
           rectangle,
           rounded corners,
           draw=black, very thick,
           text width=6.5em,
           minimum height=1em
           ,text centered
           },
    pil/.style={
           ->,
           thick,
           shorten <=2pt,
           shorten >=2pt,}
}

\newcommand{\ket}[1]{\left\vert#1\right\rangle}

\begin{document}

\title{Discrete and generalized phase space techniques in critical quantum spin chains}
\author{Zakaria Mzaouali$^{1}$}
\email{zakaria.mzaouali@um5s.net.ma}
\author{Steve Campbell$^{2}$}
\email{steve.campbell@tcd.ie}
\author{Morad El Baz$^{1,3}$}
\email{morad.elbaz@um5.ac.ma}
\affiliation{
$^1$ESMaR, Mohammed V University, Faculty of Sciences Av. Ibn Battouta, B.P. 1014,  Agdal, Rabat, Morocco. \\
$^2$School of Physics, Trinity College Dublin, Dublin 2, Ireland.\\
$^3$The Abdus Salam International Centre for Theoretical Physics, Strada Costiera 11, Miramare-Trieste, Italy.
}
\begin{abstract}
We apply the Wigner function formalism from quantum optics via two approaches, Wootters' discrete Wigner function and the generalized Wigner function, to detect quantum phase transitions in critical spin-$\tfrac{1}{2}$ systems. We develop a general formula relating the phase space techniques and the thermodynamical quantities of spin models, which we apply to single, bipartite and multi-partite systems governed by the $XY$ and the $XXZ$ models. Our approach allows us to introduce a novel way to represent, detect, and distinguish first-, second- and infinite-order quantum phase transitions. Furthermore, we show that the factorization phenomena of the $XY$ model is only directly detectable by quantities based on the square root of the bipartite reduced density matrix. We establish that phase space techniques provide a simple, experimentally promising tool in the study of many-body systems and we discuss their relation with measures of quantum correlations and quantum coherence. 
\end{abstract}
\date{\today}
\maketitle
\section{Introduction}
The use of quantum information tools in understanding many-body quantum systems continues to be a fertile line of research~\cite{DeChiaraReview, amico, sachdev}. In particular, the use of entanglement and more general forms of quantum correlation, i.e. quantum discord and coherence, to spotlight quantum phase transitions (QPTs) and extract their critical exponents has cemented the important role that such figures of merit play in unraveling the curious properties of many-body systems~\cite{niel_osb, rozario, QPT2004, qptdiscord, CampbellPRA2013, AmicoPRB, SarandyPRA2009, Werlang2010, CakmakPRB2014, qptcoherence, TonySciRep, CampbellPRB2015, CakmakPRB2016, RogersPRA2014, HofmannPRB, GiampaoloPRA, NJPCampbell, BayatPRL2017, BellIneqPRA2012, JafariPRA2017, JafariPRA2008, RulliPRA2010, Zakaria2019, cakmak_2015}. Indeed, while QPTs only strictly occur at zero-temperature, approaches based on these quantum information theoretic tools have revealed that signatures of these phenomena persist even at finite temperatures and can be rigorously studied~\cite{DeChiaraReview}. 

From quantum optics, the original continuous Wigner function, which is a quasi-probability distribution in phase space, is known to be a valuable tool in assessing the non-classical nature of systems with continuous spectra~\cite{cwigner}. The extension of the Wigner function to finite dimensional systems has been challenging, several attempts have been initiated~\cite{DWFbook}, but until recently none provided a function with the features of the continuous Wigner function. For instance, Wootters' discrete Wigner function (DWF)~\cite{wootters, wootters2, DWFJPhysA}, one of the tools at the heart of this work, can only be applied to (sub)systems having a prime dimentionality. Nevertheless, such semi-classical tools have been shown to be useful in studying the dynamic properties of many-body systems~\cite{NJPSchachenmayer, PRASchachenmayer, PRBGasenzer, QST2019, PRXSchachenmayer}. Recently the so-called generalized Wigner function (GWF) has been proposed~\cite{GWF2016} that alleviates the issues arising from other approaches for extending the use of the Wigner function and provides a complete description for any arbitrary quantum system. As we will show, there is a natural connection between the two formalisms when applied to certain settings, i.e. critical spin systems.

Recently, it has been established that the Wigner function can be used to define a bonafide measure of quantum correlations~\cite{wfrabat, wfiran}. Therefore, given the clear relationship between correlation measures and QPTs, it is natural to ask whether and how the Wigner function can be used to explore criticality. In this work we show that using phase space techniques offers a uniquely broad picture of the properties of these systems. In particular, the formalisms provide a useful tool that allows for the systematic study of single, bipartite and multipartite systems. In addition, we show that they are useful in spotlighting first, second, and continuous order QPTs, in studying ground state factorization, and both the GWF and the DWF allow us to identify which combinations of spin-spin correlation functions are relevant for characterizing the critical properties of the systems. Thus, beyond being a useful tool in the study of QPTs, we establish that such phase space techniques can provide insight into why a particular behavior may be observed for a given measure of quantum correlations across a QPT.

The remainder of the paper is organized as follows. In Sec.~\ref{sec3} we outline the DWF and the GWF formalism at the basis of our single and multi-sites analysis. We apply these techniques in Sec.~\ref{sec2} to two paradigmatic spin systems, the $XY$ model which exhibits a second-order quantum phase transition and ground-state factorization, and the $XXZ$ model which exhibits a first- and a continuous (or infinite)-order quantum phase transition. Finally, we conclude and summarize our results in Sec.~\ref{sec5}.

\section{Wigner Functions  \label{sec3}}
\subsection{Wootters' Discrete Wigner Function}
The original formulation for the Wigner function provides a phase space representation of quantum states with continuous degrees of freedom \cite{rwig1,rwig2}. For discrete systems, several methods have been developed to represent a quantum system with a finite dimensional Hilbert space  in phase space~\cite{dwig}. Among these techniques, the formalism for the discrete Wigner function (DWF) for systems with exactly $N$ (prime number) orthogonal states developed by Wootters~\cite{wootters, wootters2} provides a natural candidate for our purposes. For such systems the phase space is an $N\!\times\!N$ grid, labelled by a pair of coordinates $(x,p)$, each taking values from $0$ to $N-1$  and for each coordinate we define the usual addition and multiplication mod $N$. If the dimension of the system is $N\!=\!q^k$, with $q$ a prime and $k$ an integer greater than 1, the phase space is constructed by performing the $k$-fold Cartesian product of $q\!\times\!q$ phase spaces. Naturally, the simplest example one can consider is a system with two orthogonal states, i.e. a qubit  with $N\!=\!2$, whose discrete phase space consists of four points, while for a composite system of two qubits, i.e. $N\!=\!2^2$ the phase space is formed by 16 points, cf. Table~\ref{tab1}. Each point in the phase space is described by the discrete phase point operator, $\hat{A}(x_i,p_i)$. For a single qubit it is given by
\begin{equation}
\hat{A}(x_1,p_1)=\frac{1}{2} \Big( \openone + (-1)^{x_1} \sigma^z+(-1)^{p_1} \sigma^x+(-1)^{x_1+p_1} \sigma^y \Big),
\label{eq1}
\end{equation}
where $\sigma^i$ ($i=x,y,z$) are the usual Pauli operators. For composite systems the phase point operators are constructed from the tensor product of the phase point operators of the corresponding subsystems, i.e.  $\hat{A}(x_1\dots x_k,p_1\dots p_k)\!=\!\hat{A}(x_1,p_1)\otimes \hat{A}(x_2,p_2) \otimes \dots \otimes \hat{A}(x_k,p_k)$. Since the $\hat{A}(x_i,p_i)$'s form a complete orthogonal basis of the Hermitian $N\!\times\!N$ matrices, any density matrix can be decomposed as $\rho\!\!=\!\!\sum\limits_{(x_i,p_i)}W(x_i,p_i)\hat{A}(x_i,p_i)$, where the real-valued coefficients
\begin{equation}
W(x_i,p_i)=\frac{1}{N} \text{Tr} (\rho \hat{A}(x_i,p_i)),
\label{eq2}
\end{equation}
correspond to the DWF and $N$ is the dimension of the overall system.
\begin{table}[t]
	\begin{subtable}{.3\linewidth}
		\centering
		\captionsetup{justification=centering}
		\caption{One qubit}
		\tikzmark{t}\\
		\tikzmark{l}
		\begin{tabular}{c   c     c     c}
			& & $0$ & $1$\\[-8pt]
			& \multicolumn{1}{@{}l}{\tikzmark{x}}\\
			0 & & .&.\\
			1 & & . &.\\
		\end{tabular}
		\tikzmark{r}\\
		\tikzmark{b}
		\tikz[overlay,remember picture] \draw[-triangle 45] (x-|l) -- (x-|r) node[right] {$p_1$};
		\tikz[overlay,remember picture] \draw[-triangle 45] (t-|x) -- (b-|x) node[below] {$x_1$};
		\label{tab1a}
	\end{subtable}%
	\hfill
	\begin{subtable}{.6\linewidth}
		\centering
		\captionsetup{justification=centering}
		\caption{Two qubits}
		\tikzmark{t}\\
		\tikzmark{l}
		\begin{tabular}{c   c     c     c c c}
			& & $00$ & $01$&10&11\\[-8pt]
			& \multicolumn{1}{@{}l}{\tikzmark{x}}\\
			00 & & .&.&.&.\\
			01 & & . &.&.&.\\
			10&&.&.&.&.\\
			11&&.&.&.&.
		\end{tabular}\tikzmark{r}\\
		\tikzmark{b}
		\tikz[overlay,remember picture] \draw[-triangle 45] (x-|l) -- (x-|r) node[right] {$(p_1,p_2)$};
		\tikz[overlay,remember picture] \draw[-triangle 45] (t-|x) -- (b-|x) node[below] {$(x_1,x_2)$};
		\label{tab1b}
		\end{subtable}
	\vskip0.5cm
	\caption{Discrete phase space for (a) one qubit and (b) two qubits}
\label{tab1} 
\end{table}

\subsection{The Generalized Wigner Function}
We will also use the formalism developed by Tilma {\it et al}~\cite{GWF2016} which generalizes the Wigner function to arbitrary quantum states. Following Ref.~\cite{GWF2016}, the original Wigner function can be written in terms of the displacement $\hat{D}$ and the parity $\hat{\Pi}$ operators as
\begin{equation}
W_{\hat{\rho}}(\Omega)=\left( \frac{1}{\pi \hbar} \right)^n \text{Tr }{ \left(\hat{\rho} \hat{D}(\Omega) \hat{\Pi} \hat{D}^{\dagger}(\Omega) \right)},
\label{eq4.1}
\end{equation}
where $\hat{D}(\Omega) \hat{\Pi} \hat{D}^{\dagger}(\Omega)=\hat{\Delta}(\Omega)$ represents the kernal of the function,  $ \hat{\rho} $ is the density matrix describing the system and $\Omega$ is any full parametrization of the phase space such that $\hat{D}$ and $\hat{\Pi}$ are defined in terms of coherent states $\hat{D}(\Omega)\ket{0}=\ket{\Omega}$ and $\hat{\Pi}\ket{\Omega}=-\ket{\Omega}$. A distribution $W_{\hat{\rho}}(\Omega)$ can describe a Wigner function over a phase space parametrized by a set of $\Omega$'s, if there exists a kernel $\hat{\Delta}(\Omega)$ that generates $W_{\hat{\rho}}(\Omega)$ according to the Weyl rule
\begin{equation}
	W_{\hat{\rho}}(\Omega)=\text{Tr} \left( \hat{\rho} \hat{\Delta}(\Omega) \right),
	\label{weyl}
\end{equation}
and, as stated in~\cite{GWF2016}, also satisfy the following Stratonovich-Weyl correspondences \\
\begin{enumerate}
	\item We can fully reconstruct $\hat{\rho}$ from $W_{\hat{\rho}}(\Omega)$ and vice versa, via the mapping $W_{\hat{\rho}}(\Omega)=\text{Tr} \left( \hat{\rho} \hat{\Delta}(\Omega) \right)$ and $\hat{\rho}=\int_{\Omega} W_{\hat{\rho}} \hat{\Delta}(\Omega) d\Omega$.
	\item $W_{\hat{\rho}}$ is always real and normalises to unity.
	\item If $\hat{\rho}$ is invariant under global unitary operations then so is $W_{\hat{\rho}}$.
	\item The overlap between states, defined by the definite integral $\int_{\Omega} W_{\hat{\rho}^{\prime}} W_{\hat{\rho}^{\prime\prime}} d\Omega=\text{Tr} \left( \hat{\rho}^{\prime} \hat{\rho}^{\prime\prime} \right)$, exists and is considered a unique property of the Wigner function.
\end{enumerate}
An extension of Eq.~\eqref{eq4.1} to  finite-dimensional systems requires the construction of a kernel $\hat{\Delta}(\Omega)$ that reflects the symmetries of the system at hand. For a qubit, Tilma \textit{et al}~\cite{GWF2016} argued that the parity operator $\hat{\Pi}$ has analogous properties to $\hat{\sigma}_z$ which rotates the qubit by $\pi$ about the $z$-axis of the Bloch sphere in the Pauli representation, while the SU(2) rotation operator $\hat{U}(\theta,\varphi,\phi)=e^{i\hat{\sigma}_z\varphi}e^{i\hat{\sigma}_y\theta}e^{i\hat{\sigma}_z\phi}$ is equivalent to the displacement operator $\hat{D}$ in that $\hat{U}(\theta,\varphi,\phi)$ displaces the two level quantum state along the surface of the Bloch sphere. This line of thought leads to the following kernel for a qubit
\begin{equation}
\hat{\Delta}(\theta,\varphi)=\left[\hat{U} \hat{\Pi} \left( \hat{U} \right)^{\dagger}  \right],
\label{eq4.2}
\end{equation}
where $\theta\!\in\![0,\frac{\pi}{2}]$ and $\varphi\!\in\![0,2\pi]$ parameterize the representation in phase space and $\hat{\Pi}\!=\! \frac{1}{2} \left( \openone\!-\!\sqrt{3} \hat{\sigma}_z \right)$ is a Hermitian operator. Due to the commutation of $\hat{\sigma}_z$ with $\hat{\Pi}$,  $\phi$ does not contribute in the function. The generalization to a composite system of qubits is straightforward by performing the tensor product
\begin{equation}
	\hat{\Delta}(\theta,\varphi)=\bigotimes_i^N \hat{U} \hat{\Pi} \left( \hat{U} \right)^{\dagger}.
\end{equation}
The choice of the kernel  $\hat{\Delta}(\Omega)$ and the set of parameters $\Omega$ is not unique to define the Wigner function.

\section{Application to spin models}
\label{sec2}
In this section we apply the Wigner function formalism to two physical models of interest, the $ XY $ model and the $ XXZ $ chain both of which can be described by a real and $\mathbb{Z}_2$ symmetric Hamiltonian $H$. As they exhibit rich quantum critical behavior, quantum spin chains are the most natural candidate to investigate how phase space methods can explore criticality. 

\subsection{The $XY$ model}
The Hamiltonian of the spin$-\frac{1}{2}$ anisotropic $XY$ model with periodic boundary conditions, is given by
\begin{equation}
	\mathcal{H}_{XY}\!=\!-\sum_{i=0}^{N-1} \left[ \frac{\lambda}{2} \bigg\{ \left(1\!+\!\gamma\right)\sigma_i^x\sigma_{i+1}^x+\left(1\!-\!\gamma\right)\sigma_i^y\sigma_{i+1}^y \bigg\}\!+\!\sigma_i^z \right],
	\label{xy}
\end{equation}
where $\lambda$ is the coupling strength, $\gamma\!\!\in\!\![0,1]$ represents the anisotropy parameter, $N$ is the number of spins and $\sigma^{x,y,z}_i$ are the usual Pauli matrices. The $XY$ model is an integrable model and can be diagonalized through a Jordan-Wigner mapping followed by a Bogoliubov transformation~\cite{BarouchI, BarouchII}. In addition to the second order quantum phase transition (2QPT) occurring at $\lambda_c\!\!=\!\!1$ for $0\!<\!\gamma\!<\!1$, this model exhibits a non-trivial factorization line, where the ground state of the model becomes completely factorized
\begin{equation}
	\lambda_f=\frac{1}{\sqrt{1\!-\!\gamma^2}}
\end{equation}
and is understood as an entanglement transition which is characterized by an energy level degeneracy~\cite{GiorgiPRB, CampbellPRA2013, GiampaoloPRA, AmicoPRB}.

\subsubsection{Single site}
\begin{figure}[t]
	\includegraphics[width=0.5\columnwidth]{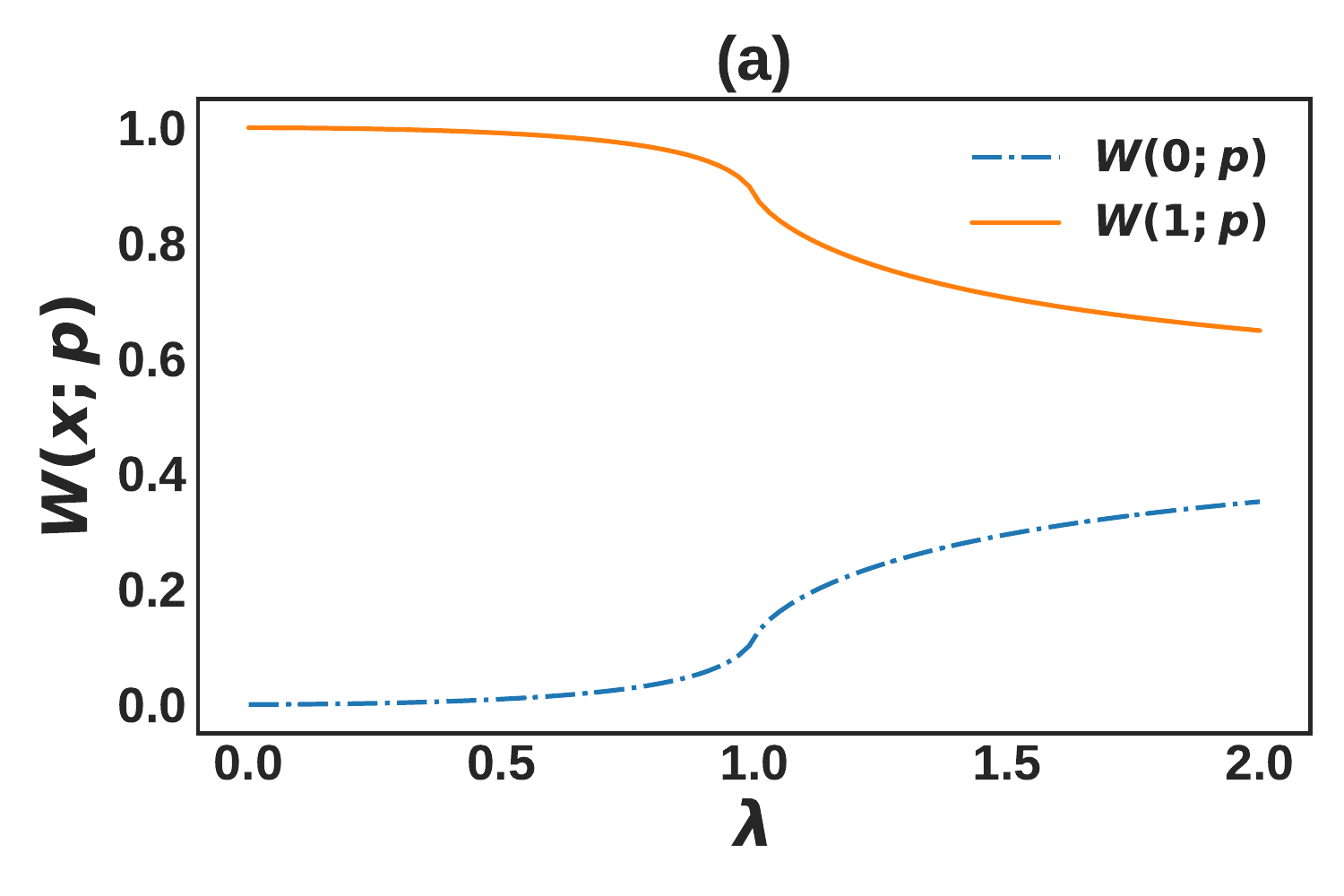}\includegraphics[width=0.5\columnwidth]{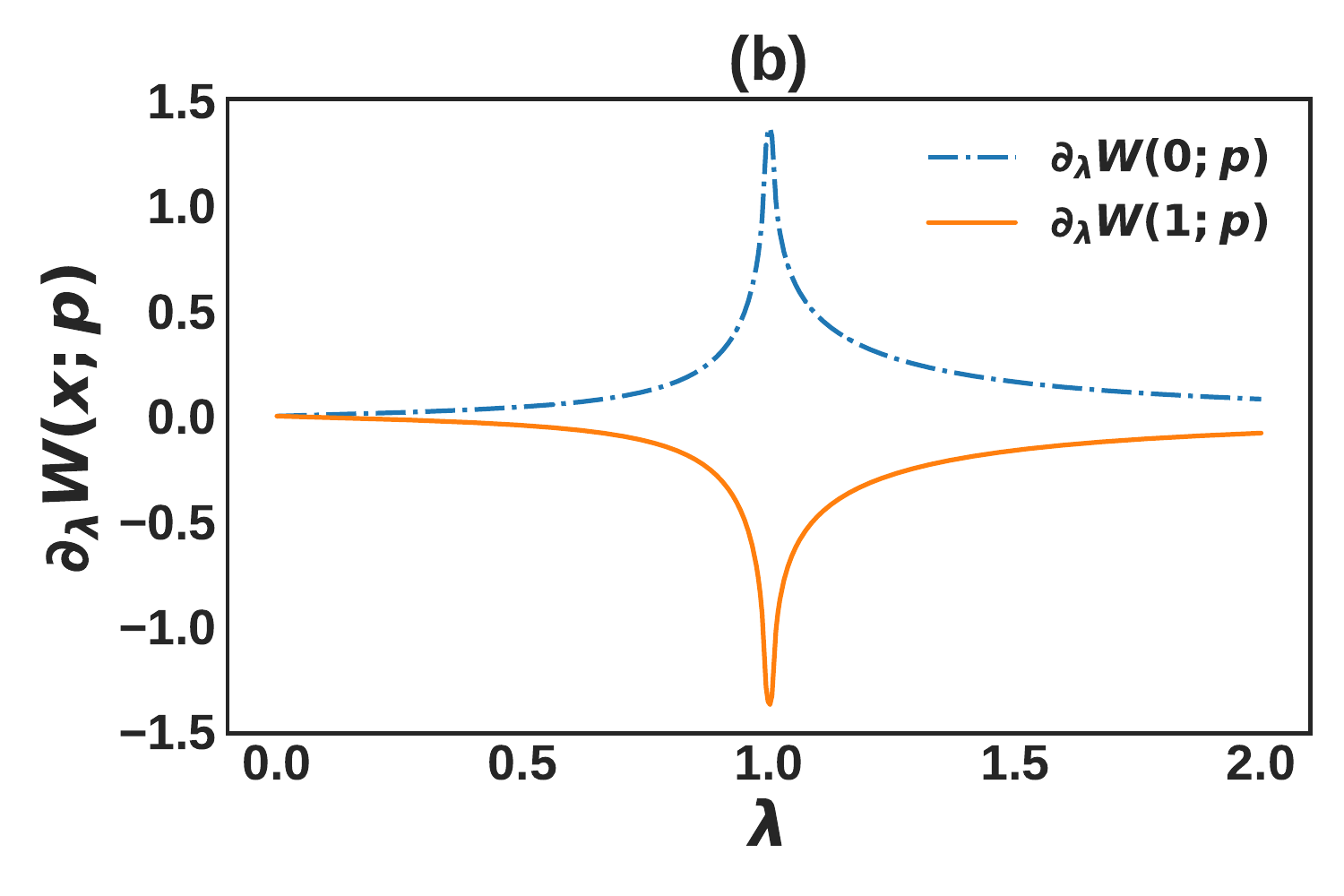}\\
	\includegraphics[width=0.5\columnwidth]{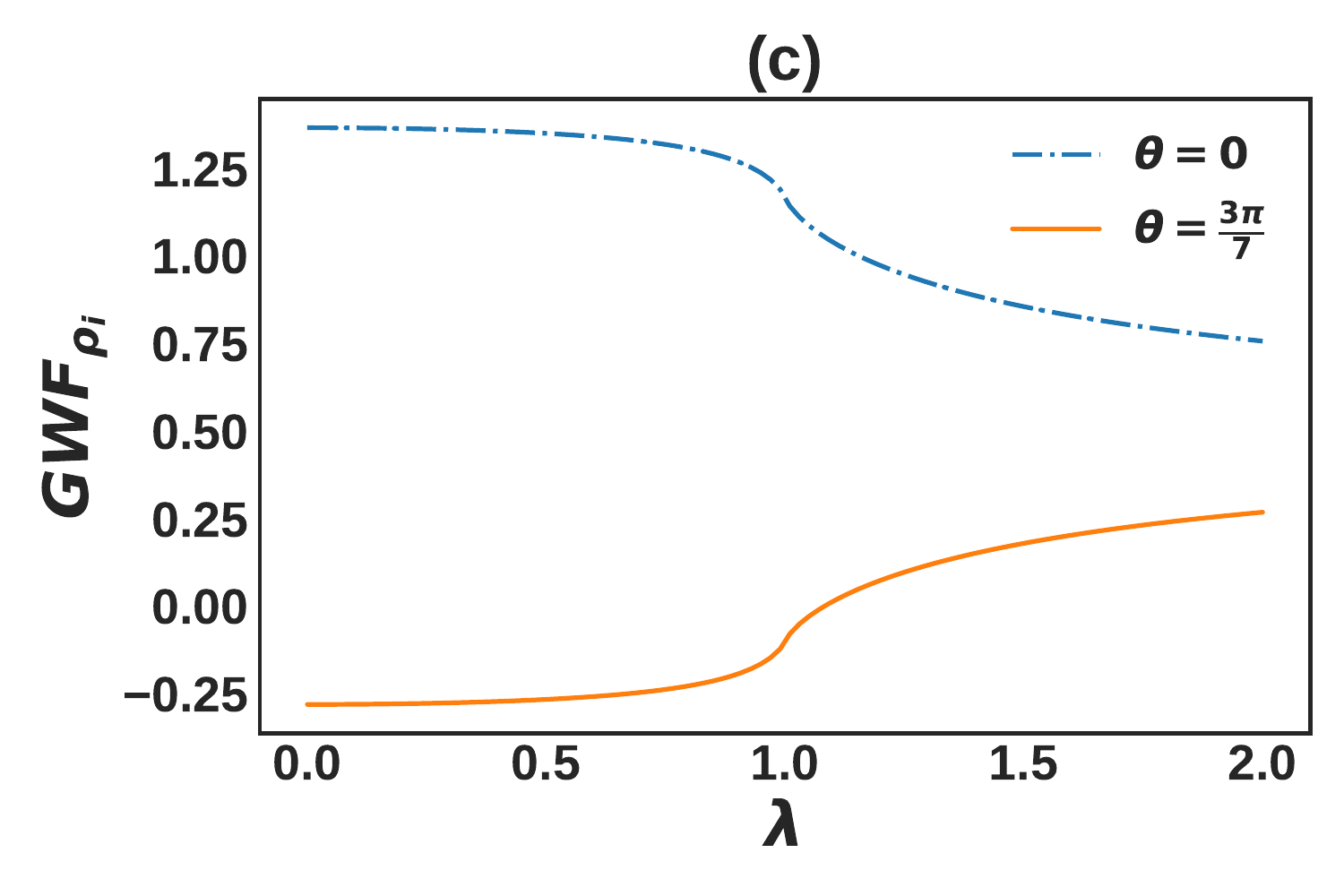}\includegraphics[width=0.5\columnwidth]{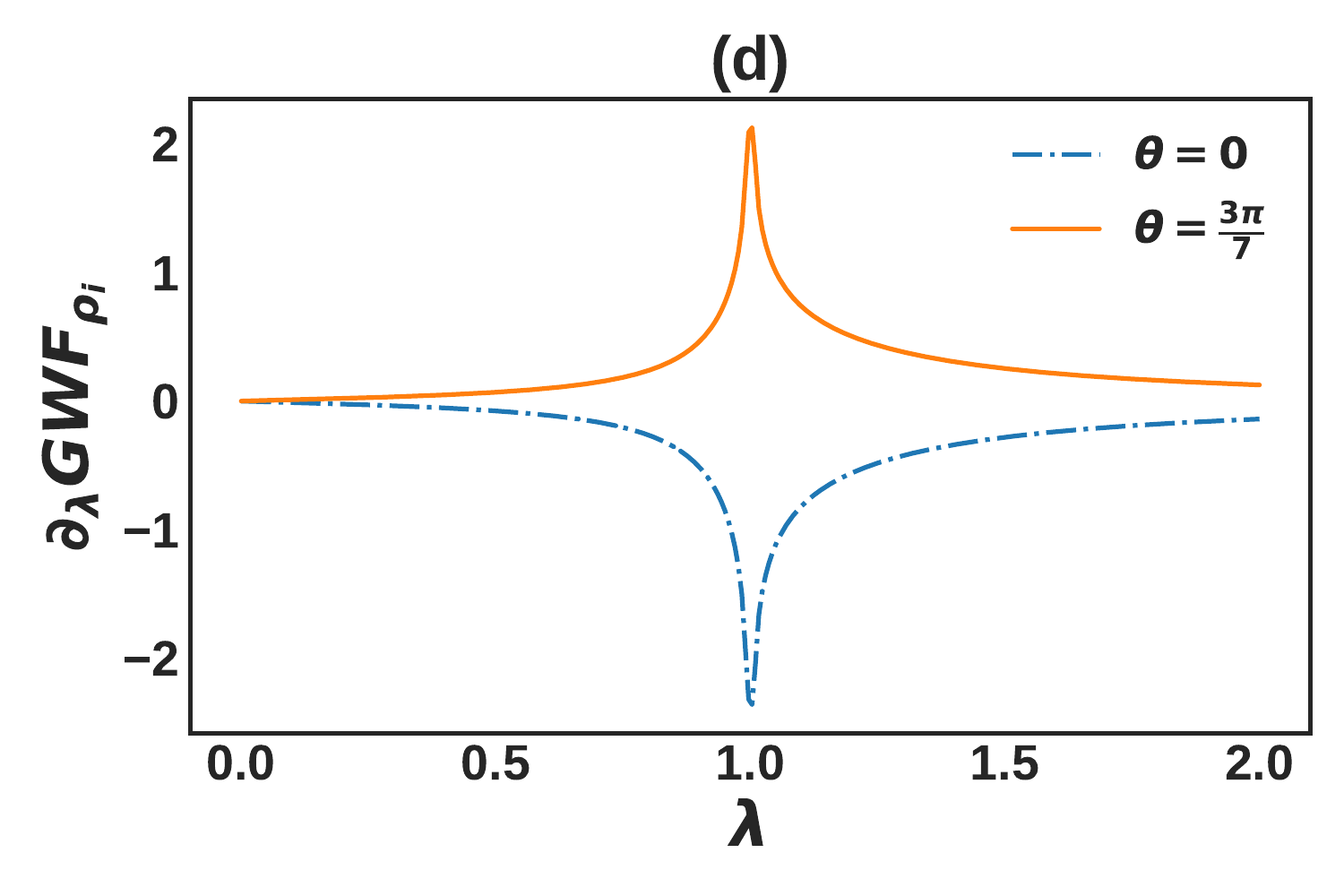}\\
	\caption{(a) Discrete Wigner function and (b) its first derivative with respect to $\lambda$ for the single site $ XY $ model, Eq.~\eqref{xy}. The two distinct behaviors of $W(x;p)$ and $\partial_{\lambda} W(x;p)$ correspond to the appropriate phase space points indicated in Table~\ref{tab_ss1}. Panels (c) and (d) show the behavior of the maximum $(\theta=0)$, the minimum $\left(\theta \sim 3\pi/7\right)$ of the GWF (c) and their first derivative with respect to $\lambda$ (d), for the $XY$ model. In all plots $\gamma\!=\!0.5$.}
	\label{fig_ssXY}
\end{figure}

We start by considering a single site taken by performing the partial trace over all the other sites of an infinite chain. The reduced density matrix $\rho_i$ can be expressed as
	\begin{equation}
		\rho_i=\frac{1}{2} \sum_{\alpha=0}^{3} \langle \sigma^\alpha \rangle \sigma_i^\alpha,
		\label{sdm}
	\end{equation}
plugging Eq.~\eqref{sdm} in Eq.~\eqref{eq2} and taking into account the reality of the density matrix and the $\mathbb{Z}_2$ symmetry of quantum spin$-\frac{1}{2}$ chains, the DWF for one site takes the form
\begin{equation}
	W(x_1,p_1)=\frac{1}{2} \Big(
	 1+(-1)^{x_1}\langle \sigma^z \rangle \Big).
	 \label{dwf_single_site} 
\end{equation}
Due to the $\mathbb{Z}_2$ symmetry, the single site DWF only depends on $x_1$ and thus the DWF in this case consists of two distinct behaviors as depicted in Fig.~\ref{fig_ssXY}(a) and (b) and Table.~\ref{tab_ss1}. Choosing $\gamma\!\!=\!\!0.5$ we see the concavity of the DWF changes after crossing the critical point, $\lambda_c\!\!=\!\!1$, which is further reflected by a divergence at $\lambda_c$ in the first derivative of the DWF with respect to $\lambda$. For this value of $\gamma$ the factorization point at $\lambda_f\!\sim\!1.1547$ and we find that the single site DWF shows no signatures of this phenomenon, which is to be expected since the reduced density matrix depends only on the magnetization and contains no information about correlations within the chain.

Turning our attention to the single site GWF, plugging the reduced density matrix Eq.~\eqref{sdm} in the Weyl rule Eq.~\eqref{weyl} we find
\begin{equation}
	\text{GWF}_{\rho_i}(\theta)\!=\!\frac{1}{2} \left(1-\sqrt{3}\cos(2\theta)\langle \sigma_z \rangle \right). 
	\label{gwf_ss}
\end{equation}
\begin{table}[t]
		\centering
		\tikzmark{t}\\
		\tikzmark{l}
		\begin{tabular}{c   c     c     c}
			& & $0$ & $1$\\[-8pt]
			& \multicolumn{1}{@{}l}{\tikzmark{x}}\\
			0 & & \begingroup \color{blue!55} \textbf{\textemdash}\endgroup&\begingroup \color{blue!55} \textbf{\textemdash} \endgroup\\
			1 & & \begingroup \color{orange!55} \textbf{-.-} \endgroup & \begingroup \color{orange!55} \textbf{-.-} \endgroup\\
		\end{tabular}
		\tikzmark{r}\\
		\tikzmark{b}
		\tikz[overlay,remember picture] \draw[-triangle 45] (x-|l) -- (x-|r) node[right] {$p_1$};
		\tikz[overlay,remember picture] \draw[-triangle 45] (t-|x) -- (b-|x) node[below] {$x_1$};
		\vskip0.5cm
		\caption{Discrete phase space for the single site $XY$ model. Each symbol corresponds to a particular curve shown in Fig.~\ref{fig_ssXY}. }
		\label{tab_ss1}
\end{table}
The single site GWF is insensitive to the angle $\varphi$ which is due to the $\mathbb{Z}_2$ symmetry. It is evident that Eq.~\eqref{dwf_single_site} corresponds to a particular choice of angle $\theta$ in the GWF, Eq.~\eqref{gwf_ss}. In Fig.~\ref{fig_ssXY} we focus on two limits of Eq.~\eqref{gwf_ss} by fixing the value of the parameter $\theta$. For $\theta\!=\!0$ ($\theta\!\sim\!3\pi/7$) Eq.~\eqref{gwf_ss} is maximized (minimized). In Fig.~\ref{fig_ssXY}(c) we see an abrupt change in both limits after crossing the critical point $\lambda_c\!=\!1$ which is reflected in the first derivative of the GWF with respect to $\lambda$ [panel (d)] by a divergence at the critical point $\lambda_c$. However, similarly to the analysis of the single site DWF, no sign of the factorization point manifests in the GWF. Due to the simple form of the single site density matrix, we find that both Eqs.~\eqref{dwf_single_site} and \eqref{gwf_ss} are directly related to the $\sigma_x$ coherence measures studied in Ref.~\cite{CakmakPRB2014}. Furthermore, this is consistent with the fact that the presence of the interference terms in the original continuous Wigner function of a given quantum state indicates quantum coherence within the state.

\subsubsection{Two site}
We extend our analysis to the case of a system of two sites $i$ and $j$ of an infinite quantum chain, with $i\!<\!j$ separated by some lattice spacing $m\!=\!j-i$. The reduced density matrix can be expressed as
\begin{equation}
\rho_{i,i+m}=\frac{1}{4}\sum_{\alpha,\beta=0}^3 p_{\alpha \beta}  \sigma_i^{\alpha}\otimes\sigma_{i+m}^{\beta},
\label{rho_ij}
\end{equation}
where $p_{\alpha \beta}=\langle\sigma_i^{\alpha}\sigma_{i+m}^{\beta}\rangle$ are the spin-spin correlation functions, ($\alpha,\beta)\!=\!0, \!1, \!2, \!3$. Substituting Eq.~\eqref{eq1} and Eq.~\eqref{rho_ij} into Eq.~\eqref{eq2} after some manipulation we find the two-site DWF can be concisely expressed as
\begin{equation}
\begin{split}
W_{\rho_{ij}}(x_1,x_2;p_1,p_2)=\frac{1}{16} \Big( 1+ \big[(-1)^{x_1}+(-1)^{x_2}\big]\langle \sigma^z \rangle\\+ (-1)^{p_1+p_2} \langle \sigma^x_i \sigma^x_{i+m} \rangle + (-1)^{x_1+x_2}\langle \sigma^z_i \sigma^z_{i+m} \rangle\\+ (-1)^{x_1+x_2+p_1+p_2} \langle \sigma^y_i \sigma^y_{i+m} \rangle \Big).
\end{split}
\label{eq4}
\end{equation}
On inspection it is evident that the DWF for a given choice of $(x_i,p_i)$ involves contributions from the various spin-spin correlation functions as well as the magnetization, which are central to spotlighting critical behavior. An advantage of Eq.~\eqref{eq4} is that it allows for a panoramic view of the properties of the system. In particular, evaluating the various DWF allows to focus on contributions that are relevant in exhibiting the critical behavior. Since a given correlation measure will often depend on only specific spin-spin correlation functions, evaluating Eq.~\eqref{eq4} also allows a window into understanding the behavior of measures of quantum correlations across QPTs. A further advantage of Eq.~\eqref{eq4} is that any given DWF is experimentally accessible~\cite{expdwf2}. It is worth emphasizing that this expression is not specific to the models considered in this work, but applies to any Hamiltonian that is real and exhibits $\mathbb{Z}_2$ symmetry.

\begin{figure}[t]
	\includegraphics[width=\columnwidth]{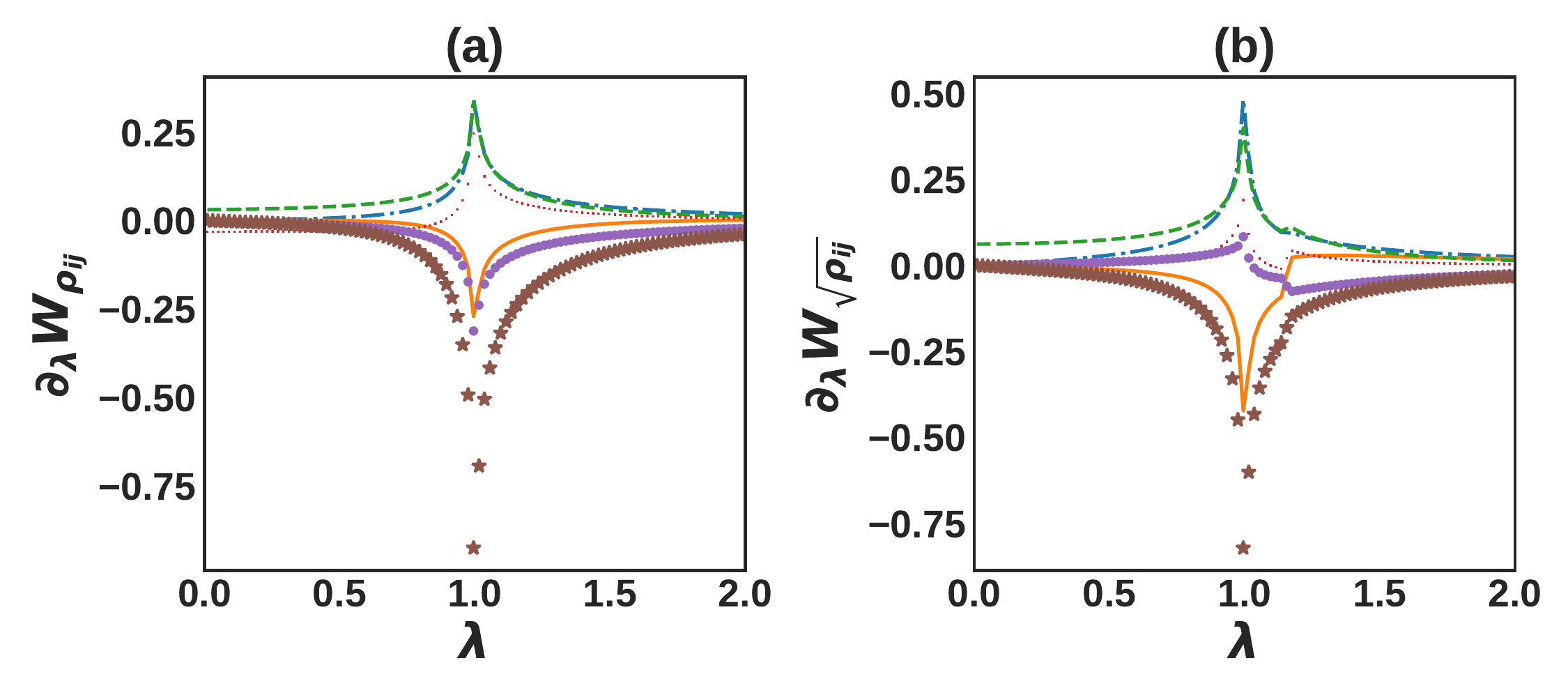}\\
	\includegraphics[width=\columnwidth]{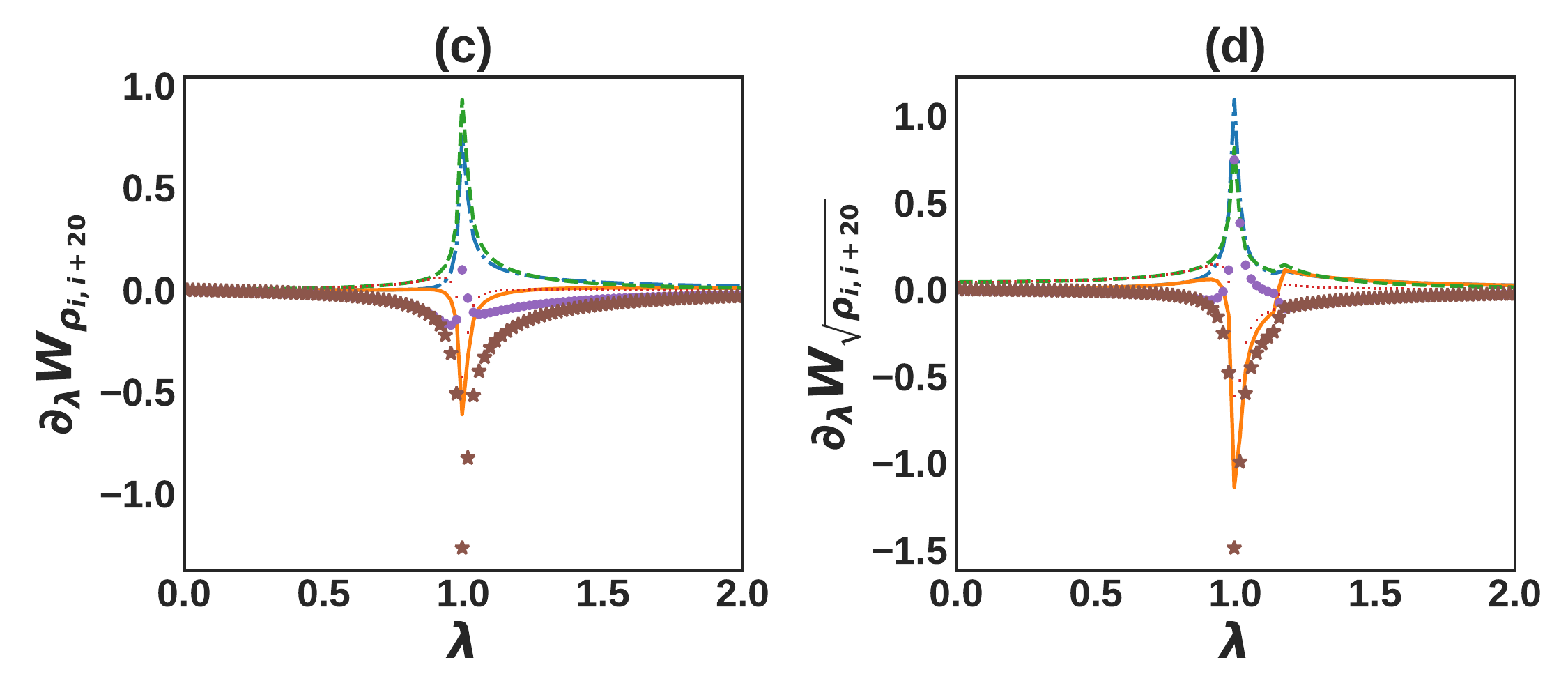}
	\caption{First derivative with respect of $\lambda$ of the discrete Wigner function of (a) $W_{\rho_{ij}}$ and (b) $W_{\sqrt{\rho_{ij}}}$ for a pair of nearest neighbor spins for the $XY$ model. The six distinct behaviors correspond to the appropriate phase space points indicated in Table~\ref{tabXY}. (c) and (d) show the same quantities for the long-range case of a pair of spins separated by 20 sites. In all panels $\gamma\!=\!0.5$}
	\label{fig_qpt_fp_xy}
\end{figure}

It is well known that various measures of bipartite quantum correlation accurately pinpoint the 2QPT~\cite{rozario, niel_osb, SarandyPRA2009, QPT2004, qptdiscord, Werlang2010, CampbellPRA2013, TonySciRep, qptcoherence, CakmakPRB2014} of this model, therefore since the DWF is constructed from combinations of correlation functions that enter into the definition of such measures, it is not surprising that we find a qualitatively similar behavior. In line with these previous studies, Fig.~\ref{fig_qpt_fp_xy}(a) shows the first derivative of the DWFs for a pair of nearest neighbor spins. We see that there are six characteristic behaviors, cf. Table~\ref{tabXY}, and all of them clearly signal the 2QPT by showing a divergence in the first derivative at the critical point. Thus, as all discrete phase space points exhibit a qualitatively similar behavior, choosing to study any one in particular is sufficient to study the QPT.

It is interesting to note that, despite being dependent on all the relevant spin-spin correlation functions, there is no evidence of ground state factorization in the behavior of $\partial_{\lambda}W_{\rho_{ij}}$. Furthermore, it was shown for some coherence measures~\cite{CakmakPRB2014}, the factorization phenomenon is connected to an inherited discontinuity at the level of $\sqrt{\rho_{ij}}$ instead of $\rho_{ij}$. Inspired by this observation, we find a consistent behavior in the DWF by calculating $\partial_{\lambda} W_{\sqrt{\rho_{ij}}}$ in Fig.~\ref{fig_qpt_fp_xy}(b). Now we find that the DWF develops a finite discontinuity at the factorization point for all six characteristic behaviors. The physical significance of this observation remains to be understood~\cite{cakmak_2015}.

\begin{table}[t]
	\begin{subtable}{\linewidth}
		\centering
		\tikzmark{t}\\
		\tikzmark{l}
		\begin{tabular}{c   c     c     c c c}
			& & $00$ & $01$&10&11\\[-8pt]
			& \multicolumn{1}{@{}l}{\tikzmark{x}}\\
			00 & & \begingroup \color{blue!55} \textbf{-.-}\endgroup&\begingroup \color{orange} \textbf{\textemdash} \endgroup&\begingroup \color{orange} \textbf{\textemdash} \endgroup&\begingroup \color{blue!55} \textbf{-.-}\endgroup\\
			01 & & \begingroup \color{green} \textbf{ - - - }\endgroup&\begingroup \color{violet} \textbf{ \textellipsis\textellipsis }\endgroup&\begingroup \color{violet} \textbf{ \textellipsis\textellipsis }\endgroup&\begingroup \color{green} \textbf{ - - - }\endgroup \\
			10 & & \begingroup \color{green} \textbf{ - - - }\endgroup&\begingroup \color{violet} \textbf{ \textellipsis\textellipsis }\endgroup&\begingroup \color{violet} \textbf{ \textellipsis\textellipsis }\endgroup&\begingroup \color{green} \textbf{ - - - }\endgroup\\
			11 & & \begingroup \color{red} \huge $,$ \endgroup&\begingroup \color{Sepia} $\bigstar$ \endgroup&\begingroup \color{Sepia} $\bigstar$ \endgroup&  \begingroup \color{red} \huge $,$ \endgroup \\
		\end{tabular}\tikzmark{r}\\
		\tikzmark{b}
		\tikz[overlay,remember picture] \draw[-triangle 45] (x-|l) -- (x-|r) node[right] {$(p_1,p_2)$};
		\tikz[overlay,remember picture] \draw[-triangle 45] (t-|x) -- (b-|x) node[below] {$(x_1,x_2)$};
	\end{subtable}
	\vskip0.5cm
	\caption{Discrete phase space of the $ XY $ model . Each symbol corresponds to a particular curve shown in Fig.~\ref{fig_qpt_fp_xy}.}
	\label{tabXY}
\end{table}

Finally we examine the long range behavior of the DWF in the $XY$ model. Fig.~\ref{fig_qpt_fp_xy}(c) and (d) depicts the first derivative of the DWF for a pair of spins $i$ and $j\!=\!i+m$ separated by $m\!=\!20$. For all phase space points, both first derivatives with respect to $\lambda$ of $W_{\rho_{i,i+20}}$ and $W_{\sqrt{\rho_{i,i+20}}}$ diverges at the critical point $\lambda_c\!=\!1$, revealing the 2QPT, while the discontinuity at the factorization point persists at long range only at the level of the first derivative of $W_{\sqrt{\rho_{i,i+20}}}$, which is consistent with the previous finding in the case of nearest neighbors.

As with the single site case, we extend the GWF formalism to the case of a system composed of two sites $i$ and $j$ with $i\!<\!j$, separated by some lattice spacing $m\!=\!j-i$. Plugging the reduced density matrix $\rho_{ij}$, Eq.~\eqref{rho_ij}, into the Weyl rule Eq.~\eqref{weyl}, the two site GWF can be expressed as
\begin{align}
&\text{GWF}_{\rho_{ij}}(\theta_i,\varphi_i,\theta_j,\varphi_j)\!=\!\frac{1}{4} \Big[ 1-\sqrt{3} \left(\cos{2\theta_i} + \cos{2\theta_j} \right) \langle \sigma^z \rangle + \nonumber\\&
3\cos{2\varphi_i}\sin{2\theta_i}\cos{2\varphi_j}\sin{2\theta_j}  \langle \sigma^x_i\sigma^x_{j} \rangle + 3\sin{2\theta_i}\sin{2\theta_j} \nonumber \\&\sin{2\varphi_i}\sin{2\varphi_j} \langle \sigma^y_i\sigma^y_{j} \rangle +
3\cos{2\theta_i}\cos{2\theta_j} \langle \sigma^z_i\sigma^z_{j} \rangle \Big].
\label{gwf_ij}
\end{align}
In line with the DWF analysis, the GWF is written in terms of the various correlation functions with the additional dependence on the set of angles $(\theta_i,\varphi_i,\theta_j,\varphi_j)$. Again we see that Eq.~\eqref{gwf_ij} is an extension of Wootters' DWF. 
\begin{figure}[t!]
	\centering
	\includegraphics[width=\columnwidth]{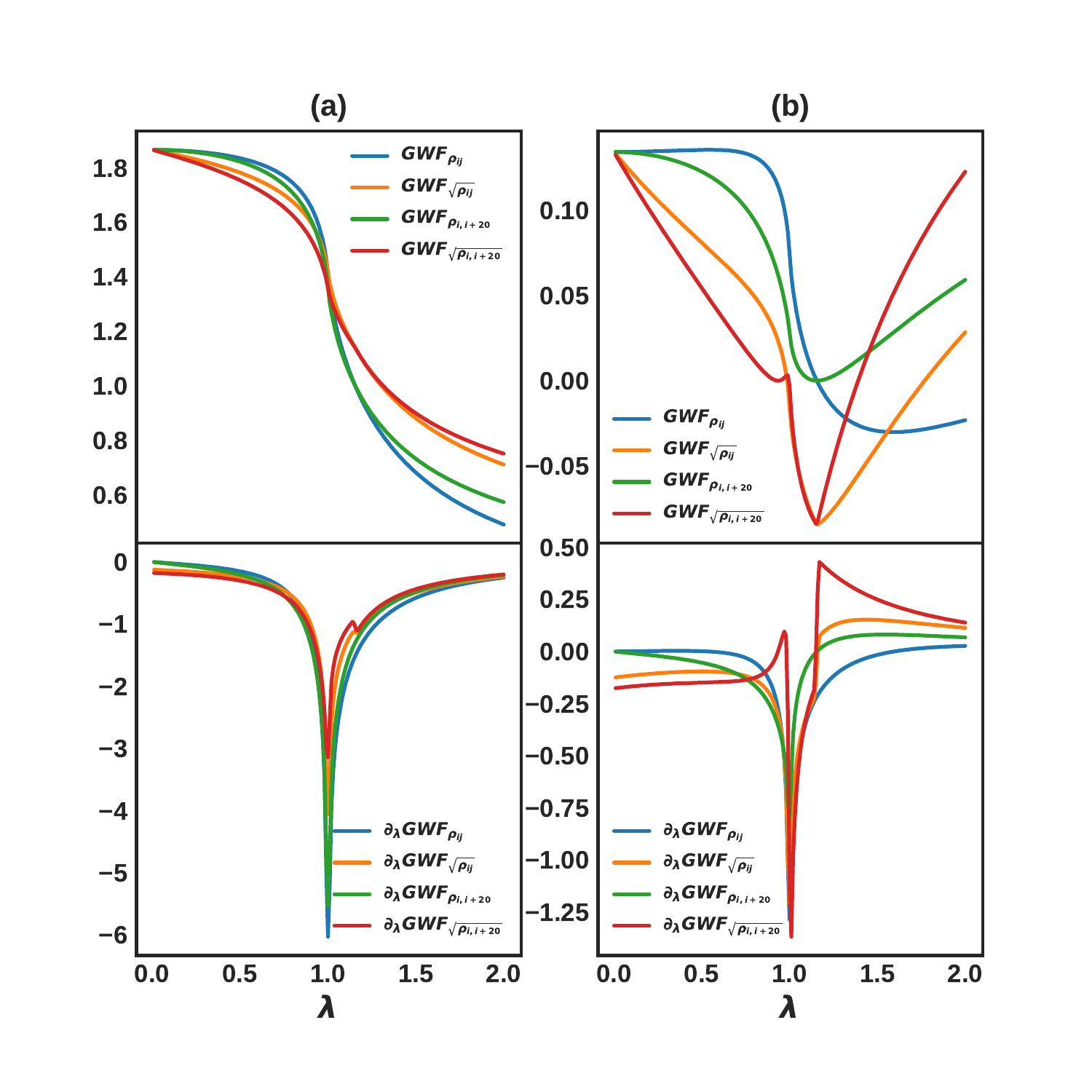}
	\caption{The two sites GWF of the $XY$ model taking $\gamma=0.5$ in Eq.~\eqref{gwf_ij} [upper panels] and its first derivative with respect to $\lambda$ [lower panels] for (a) $\theta_i\!=\!\theta_j\!=\!\pi/2$; $\varphi_i\!=\!\varphi_j\!=\!2\pi$ and (b) $ \theta_i\!=\!\theta_j\!=\!\varphi_i\!=\!\varphi_j\!=\!0  $ in the case of nearest and the $20^{\text{th}}$ neighbor using $\rho_{ij}$ and $\sqrt{\rho_{ij}}$.}
	\label{fig_gwf_qpt_fp}.
\end{figure}

Fig.~\ref{fig_gwf_qpt_fp} shows the behavior of the two site GWF for the $XY$ model with $\gamma=0.5$ for two angle configurations: $(\theta_i\!=\!\theta_j\!=\!\varphi_i\!=\!\varphi_j\!=\!0)$ and $(\theta_i\!=\!\theta_j\!=\!\pi/2 ; \varphi_i\!=\!\varphi_j\!=\!2\pi)$. Focusing on the upper panels of (a) and (b), we see an inflection point for both configurations for nearest neighbors (blue line) and $20^{\text{th}}$ neighbors (green line) at the critical point $\lambda_c\!=\!1$. As with the DWF, the factorization point can only be directly detected by examining the GWF for ${\sqrt{\rho_{ij}}}$, where a discontinuity appears in the derivative. However, the factorization phenomenon comes with an additional property: the value of the GWF {\it at the factorization point} is constant for any lattice distance $m$ which can be seen clearly for the $\text{GWF}_{\rho_{ij}}$ and $\text{GWF}_{\sqrt{\rho_{ij}}}$ in Fig.~\ref{fig_gwf_qpt_fp}. Such a behavior was first noted for the quantum discord in the same model~\cite{EPL2011, CampbellPRA2013}. This property can be understood when looking at the energy levels of finite-sizes of the $XY$ chain, where an energy-level crossing between the ground state and the first excited state take place exactly at the factorization point~\cite{CampbellPRA2013, AmicoPRB, GiorgiPRB} which forces the spin-spin correlation functions to have a constant value at any distance $m$. 

\begin{figure}[t]
	\centering
	\includegraphics[width=\columnwidth]{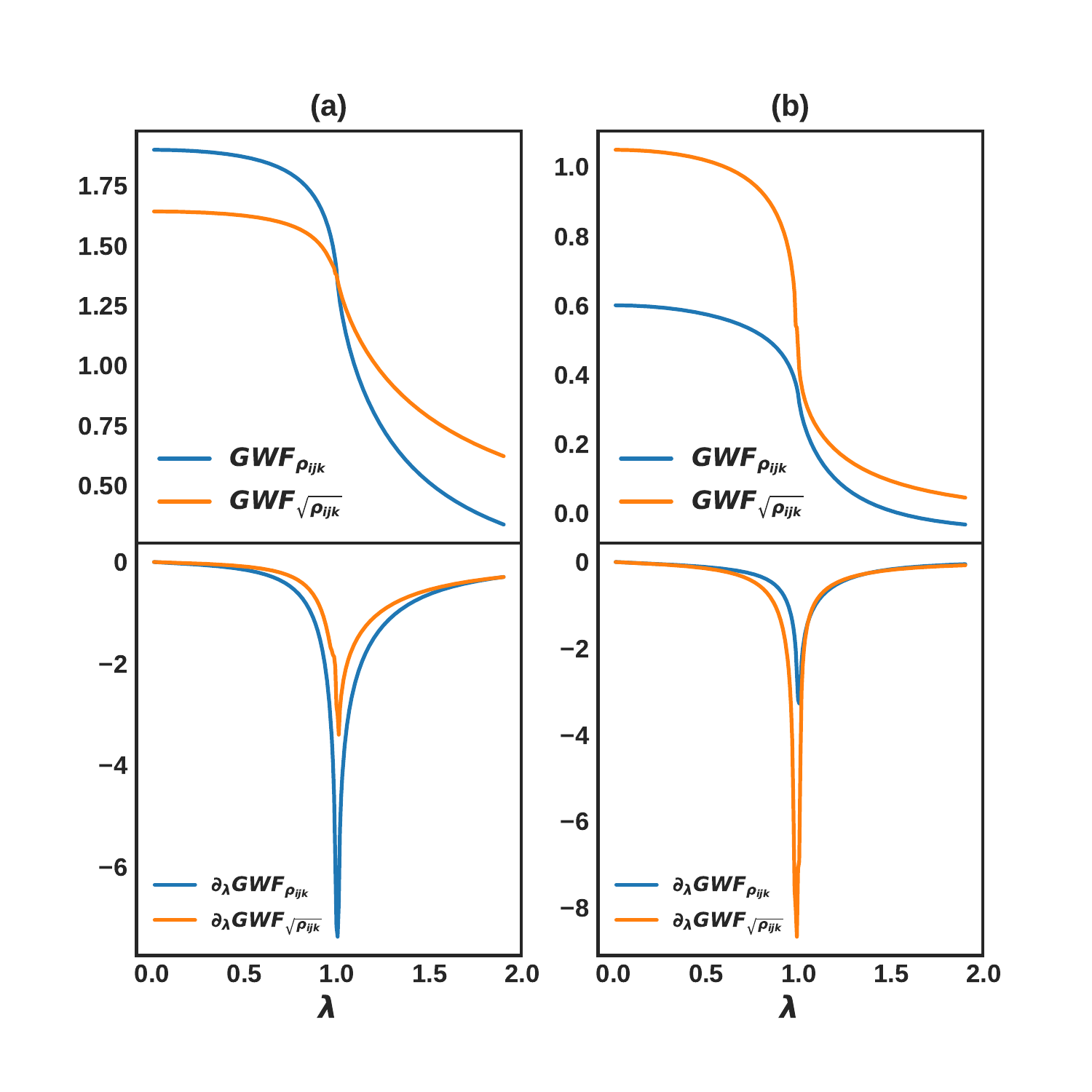}
	\caption{The three sites GWF of the $XY$ model taking $\gamma=0.5$ in Eq.~\eqref{gwf_ijk} and its first derivative with respect to $\lambda$ for (a) $\theta_i\!=\!\theta_j\!=\!\pi/2$; $\varphi_i\!=\!\varphi_j\!=\!2\pi$ and (b) $ \theta_i\!=\!\theta_j\!=\!\varphi_i\!=\!\varphi_j\!=\!0  $ in the case of nearest neighbor spins.}
	\label{fig_gwf_ijk}
\end{figure}

\begin{figure*}
	{(a)} \hskip0.7\columnwidth {(b)}\hskip0.7\columnwidth {(c)}\\
	\includegraphics[width=0.7\columnwidth]{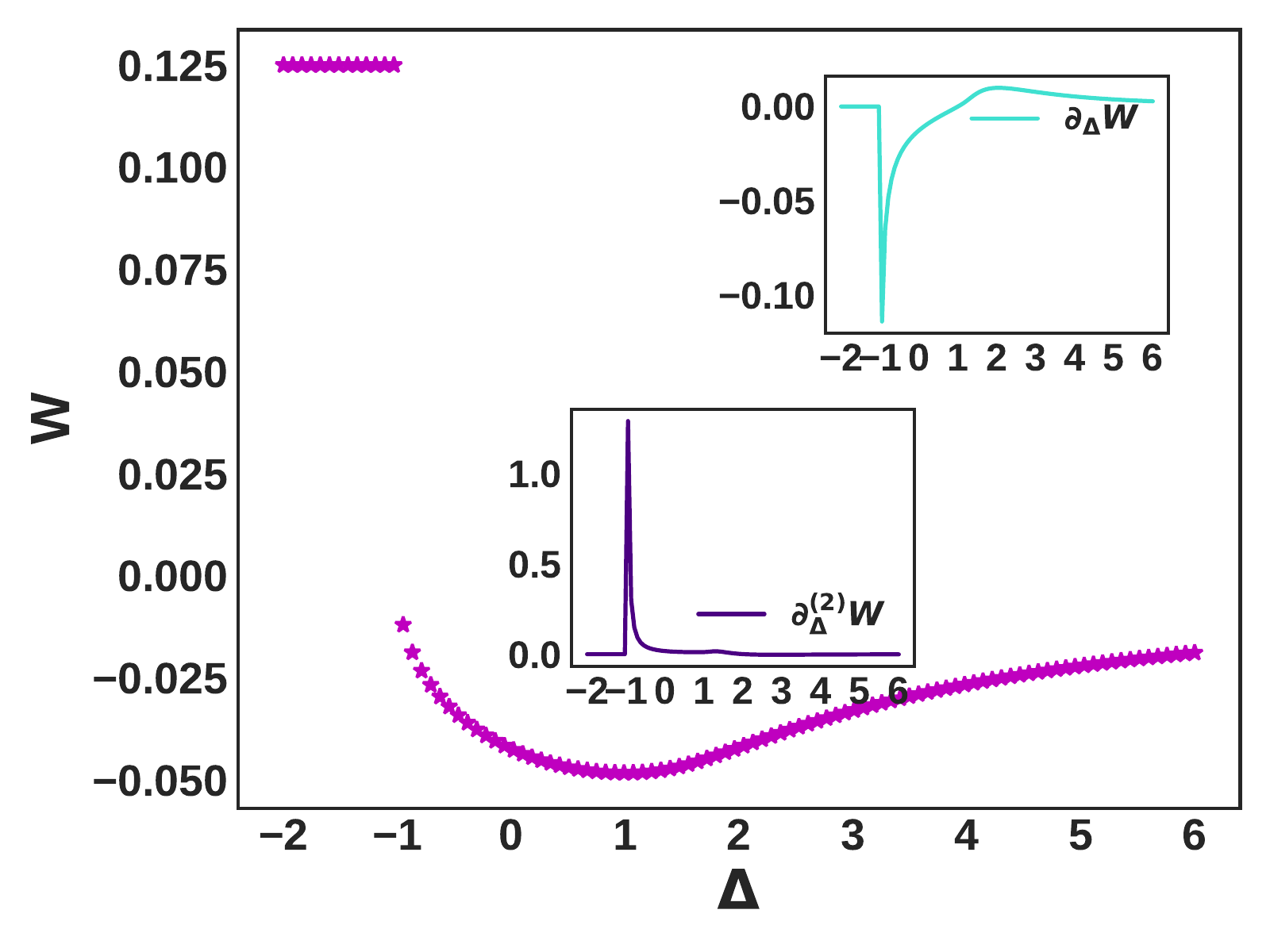}\includegraphics[width=0.7\columnwidth]{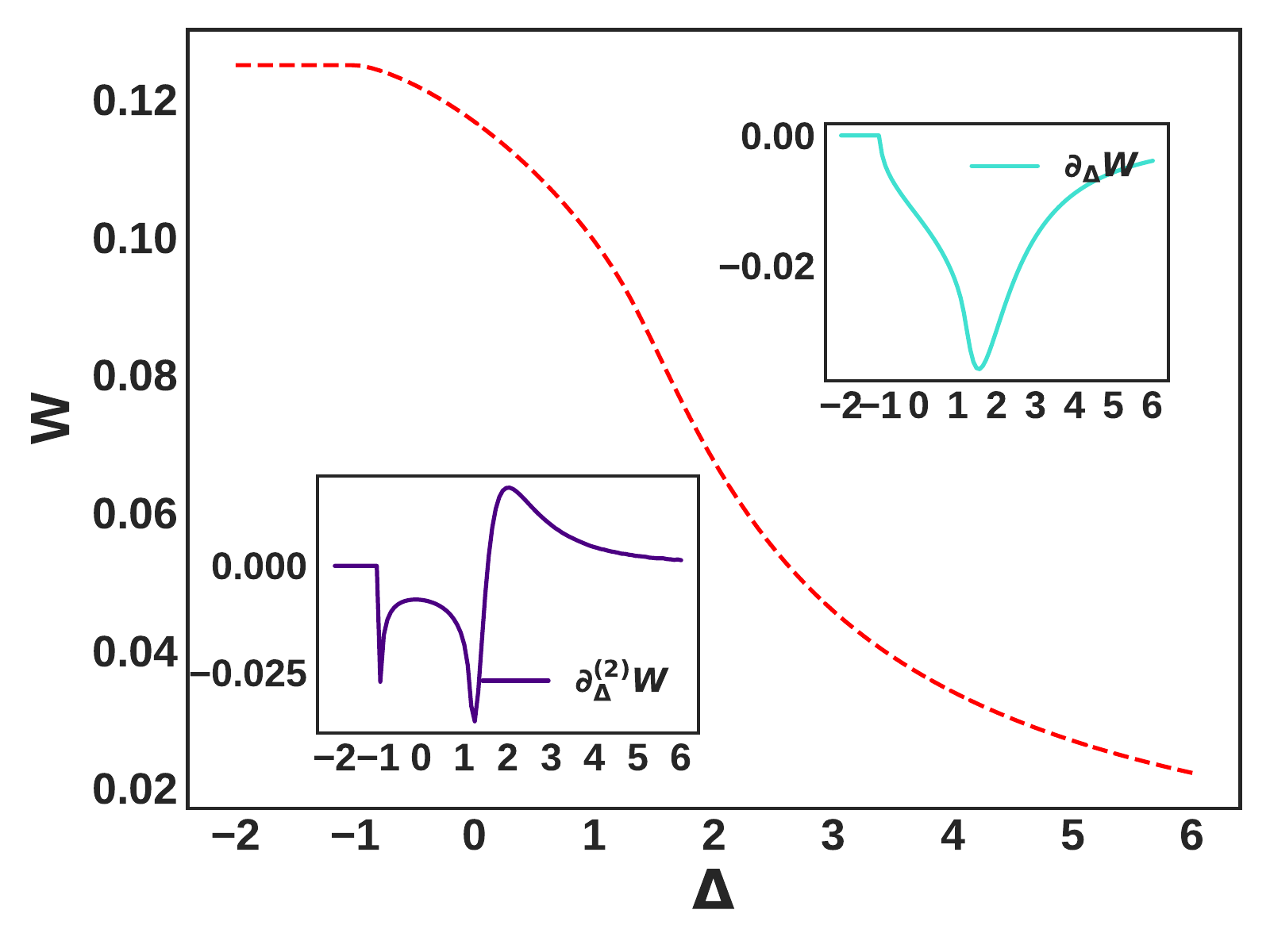}\includegraphics[width=0.7\columnwidth]{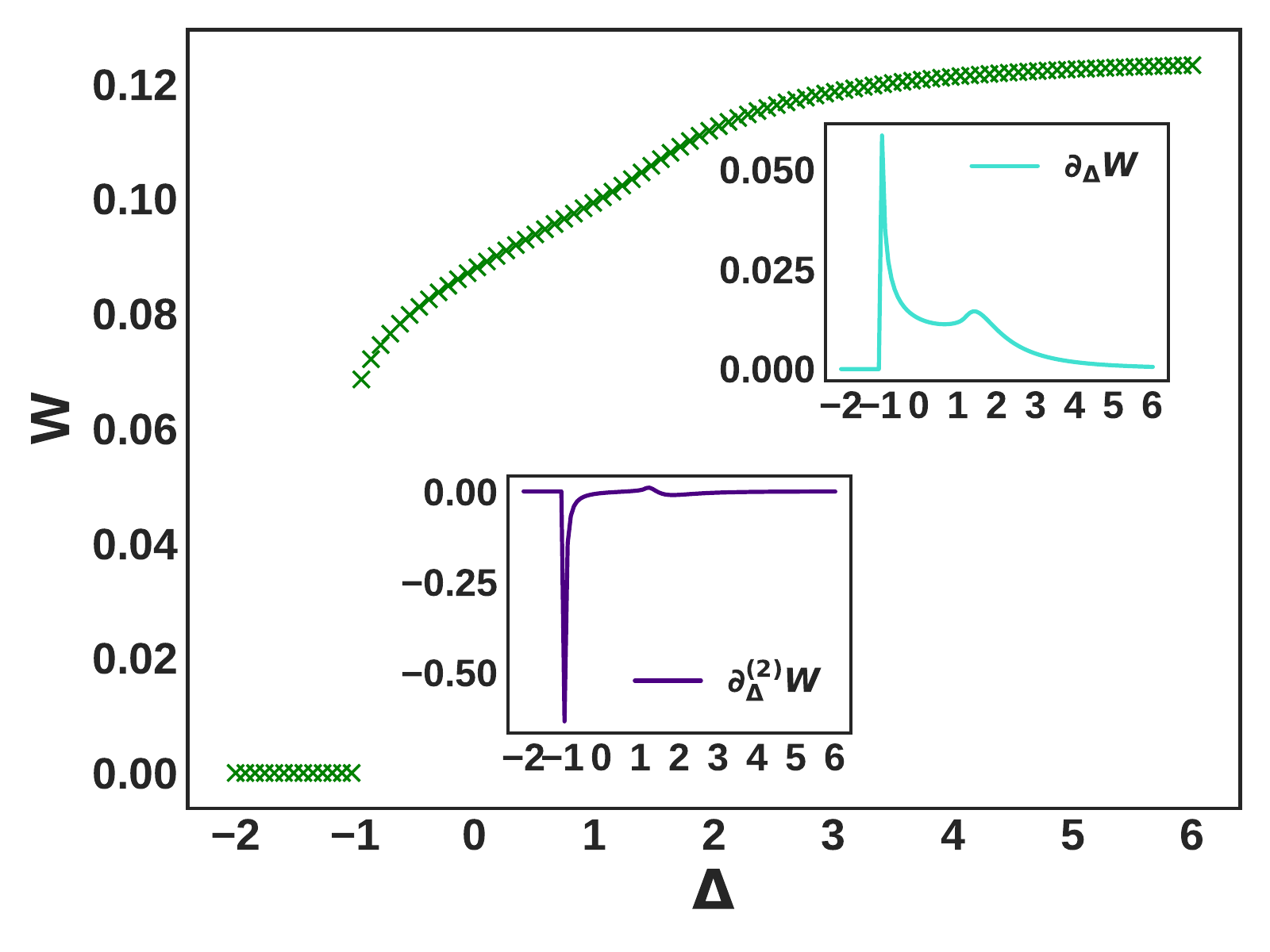}
	\caption{Discrete Wigner function for a pair of nearest neighbor spins in for the $XXZ$ model, Eq.~\eqref{eq6}. The three distinct behaviors correspond to the appropriate phase space points indicated in Table~\ref{tabxxz}.}
	\label{fig_XXZ}
\end{figure*}
\subsubsection{Multi-sites}
A significant advantage of the GWF approach is it can readily be extended to multipartite systems. Here we examine a three site system of the $XY$ model. The reduced density matrix $\rho_{ijk}$, taken by performing the partial trace over the infinite chain except the sites $(i,j,k)$, expressible as $\rho_{ijk}=\frac{1}{2^3} \sum_{\alpha,\beta,\gamma=0}^3 \langle \sigma_{i}^{\alpha} \sigma_{j}^{\beta} \sigma_{k}^{\gamma} \rangle \sigma_{i}^{\alpha} \otimes \sigma_{j}^{\beta} \otimes \sigma_{k}^{\gamma}$
and the full expression for the GWF, Eq.~\eqref{gwf_ijk} is provided in the appendix. While in principle one could also consider the DWF for this case, in general the discrete phase space will consist of 64 behaviors, making it difficult to visualize. Furthermore, as we have established from the single and two-site analyses, the DWF corresponds to particular choices for the angles entering into the GWF. In analogy with the analysis of the two sites GWF, we consider two angle configurations: $\theta_i\!=\!\theta_j\!=\!\theta_k\!=\!\pi/2;\varphi_i\!=\!\varphi_j\!=\!\varphi_k\!=\!2\pi$ and $\theta_i\!=\!\theta_j\!=\!\theta_k\!=\!\varphi_i\!=\!\varphi_j\!=\!\varphi_k\!=\!0$ of the three sites GWF. In Fig.~\ref{fig_gwf_ijk} we see that in the multipartite case continues to spotlight the second order QPT for both sets of angles, however no sign of the factorization point can be seen. The failure of a multipartite non-classicality indicator to witness the factorization point is remarkable and is at variance with the behavior of certain indicators of multipartite entanglement which vanish in the thermodynamic limit~\cite{GiampaoloPRA}.

\subsection{The $XXZ$ model}
As a second interesting candidate system, we consider the $XXZ$ model with periodic boundary conditions
\begin{equation}
\mathcal{H}_{XXZ}=\frac{1}{4}\sum_{i=1}^N \sigma_i^x\sigma_{i+1}^x+\sigma_i^y\sigma_{i+1}^y+\Delta \sigma_i^z\sigma_{i+1}^z,
\label{eq6}
\end{equation}
where $\Delta$ is the anisotropy parameter. The phase diagram is split into three regions, separated by two different QPTs. For $\Delta\!\leq\!-1$, the system is in a ferromagnetic (gapped) phase and at $\Delta\!=\!-1$ a first-order quantum phase transition (1QPT) occurs. For $-1\!<\!\Delta\!<\!1$, the system is in a gapless (Luttinger liquid) phase and at $\Delta\!=\!1$ an infinite-order continuous quantum phase transition (CQPT) occurs, known as the Kosterlitz-Thouless QPT~\cite{Kosterlitz1973}. Finally, for $\Delta\!>\!1$, the system enters the anti-ferromagnetic (gapped) phase. The equilibrium properties of this model have been well studied, and in particular various measures of bipartite quantum correlations and their behavior across the different QPTs have been explored~\cite{qptdiscord,JafariPRA2008, SarandyPRA2009, RulliPRA2010, JafariPRA2017}. While entanglement and quantum discord were shown to reveal the critical points, their qualitative behaviors were shown to be strikingly different~\cite{SarandyPRA2009}. Here, by examining the DWF and the GWF we can shed greater light on these behaviors and show that when extremization procedures are employed, features spotlighting criticality becomes more pronounced.
\begin{table}[t]
    \begin{subtable}{\linewidth}
            \centering
            \tikzmark{t}\\
            \tikzmark{l}
        \begin{tabular}{c   c     c     c c c}
        & & $00$ & $01$&10&11\\[-8pt]
        & \multicolumn{1}{@{}l}{\tikzmark{x}}\\
        00 & & \begingroup \color{magenta}$ \bigstar$ \endgroup&\begingroup \color{red} \textbf{ - - -  } \endgroup&\begingroup \color{red} \textbf{ - - -  } \endgroup&\begingroup \color{magenta} $ \bigstar$ \endgroup\\
        01 & & \begingroup \color{green} $\times$ \endgroup&\begingroup \color{green} $\times$ \endgroup&\begingroup \color{green} $\times$ \endgroup&\begingroup \color{green} $\times$ \endgroup \\
        10 & & \begingroup \color{green} $\times$ \endgroup&\begingroup \color{green} $\times$ \endgroup&\begingroup \color{green} $\times$ \endgroup&\begingroup \color{green} $\times$ \endgroup\\
        11 & & \begingroup \color{magenta} $ \bigstar$ \endgroup&\begingroup \color{red} \textbf{ - - -  } \endgroup&\begingroup \color{red} \textbf{ - - -  } \endgroup&\begingroup \color{magenta}$ \bigstar$ \endgroup \\
        \end{tabular}\tikzmark{r}\\
        \tikzmark{b}
        \tikz[overlay,remember picture] \draw[-triangle 45] (x-|l) -- (x-|r) node[right] {$(p_1,p_2)$};
        \tikz[overlay,remember picture] \draw[-triangle 45] (t-|x) -- (b-|x) node[below] {$(x_1,x_2)$};
        \end{subtable}
        \vskip0.5cm
 \caption{Discrete phase space of the $XXZ$ model. Each symbol corresponds to one of the three characteristic behaviors shown in Fig.~\ref{fig_XXZ}.}
        \label{tabxxz}
\end{table}

Due to the form of Eq.~\eqref{eq6} we find that no relevant information about the critical properties of the system can be revealed by studying only the single site density matrix. This is simply due to the fact that the single site density matrix depends only on $\langle \sigma^z \rangle$, which is constant for the $XXZ$ model. Therefore, for the remainder we will focus on the two site setting.

For the $XXZ$ model we find that the two-site DWF calculated following Eq.~\eqref{eq4} exhibits three distinct behaviors shown in Fig.~\ref{fig_XXZ} and Table~\ref{tabxxz} as a function of $\Delta$, and their corresponding first and second derivatives for nearest-neighboring sites. Let us first consider the behavior of the DWF at the corners of the discrete phase space i.e. (00,00), (00,11), (11,00) and (11,11) [cf. Fig.~\ref{fig_XXZ} (a)]. We see that the DWF is discontinuous at the 1QPT $\Delta\!\!=\!\!-1$ while it reaches a minimum at the CQPT at $\Delta\!\!=\!\!1$, after which the DWF approaches zero with increasing anisotropy. This behavior is qualitatively identical to that of  the entanglement measured via concurrence which in this case is simply $2 |\langle \sigma_i^x \sigma_j^x \rangle|$. The relationship is evident due to the fact that the DWF at these points depends on both $\langle \sigma_i^x \sigma_j^x \rangle$ and $\langle \sigma_i^z \sigma_j^z \rangle$. It is interesting that by direct calculation we confirm that the negativity of the DWF coincides with the presence of entanglement in the state, inline with a negative behavior of the continuous Wigner function implying genuine non-classicality of the state~\cite{wfrabat, wfiran}.

The second significant behavior is located at phase space points (00,01), (00,10), (11,01), and (11,10) shown in Fig.~\ref{fig_XXZ} (b) where, in contrast with the previous cases, signatures of the critical points are less evident immediately in the behavior of the DWF. For $\Delta\!\!<\!\!-1$ these functions are constant and exhibit a sudden change at the 1QPT. On inspection we can see a point of inflection around $\Delta\!=\!1.5$. Looking at the first derivative of the DWF we see that it presents an amplitude bump at $\Delta=1.5$, and the second derivative is divergent at $\Delta\!=\!-1$ and around $\Delta\!=\!1$.  The more peculiar behavior seen in this DWF is due to the destructive interference at these phase space points between the two terms that control the DWF which are $1+\langle \sigma_i^z \sigma_j^z \rangle$ and $-2\langle \sigma_i^x \sigma_j^x \rangle$, and the inflection point seen arises from a sudden change in the concavity of $-2\langle \sigma_i^x \sigma_j^x \rangle$. Thus, unlike in the $XY$ model where all DWFs readily witness the 2QPT, the DWF in these four points can only easily signal the 1QPT exactly, while for the CQPT it shows only some anomalies around $\Delta\!\!=\!\!1$. However, we will revisit this behavior in the context of extremization procedures shortly.

Finally we consider the remaining eight phase space points, Fig.~\ref{fig_XXZ} (c). Here the DWF depends solely on a single term, $1-\langle \sigma_i^z \sigma_j^z \rangle$, and owing to the fact that spin-spin correlation functions are discontinuous at $\Delta\!\!=\!\!-1$ and that on their own they fail at revealing the CQPT at $\Delta\!=\!1$ the DWF at these points inherits these properties from the $\langle \sigma_i^z \sigma_j^z \rangle$ contribution which explains why the DWF is discontinuous and its derivatives are divergent at $\Delta\!=\!-1$, while it does not show any special behavior at the CQPT $\Delta\!=\!1$.

\begin{figure}[t]
{(a)} \hskip0.5\columnwidth {(b)}
\includegraphics[width=0.5\columnwidth]{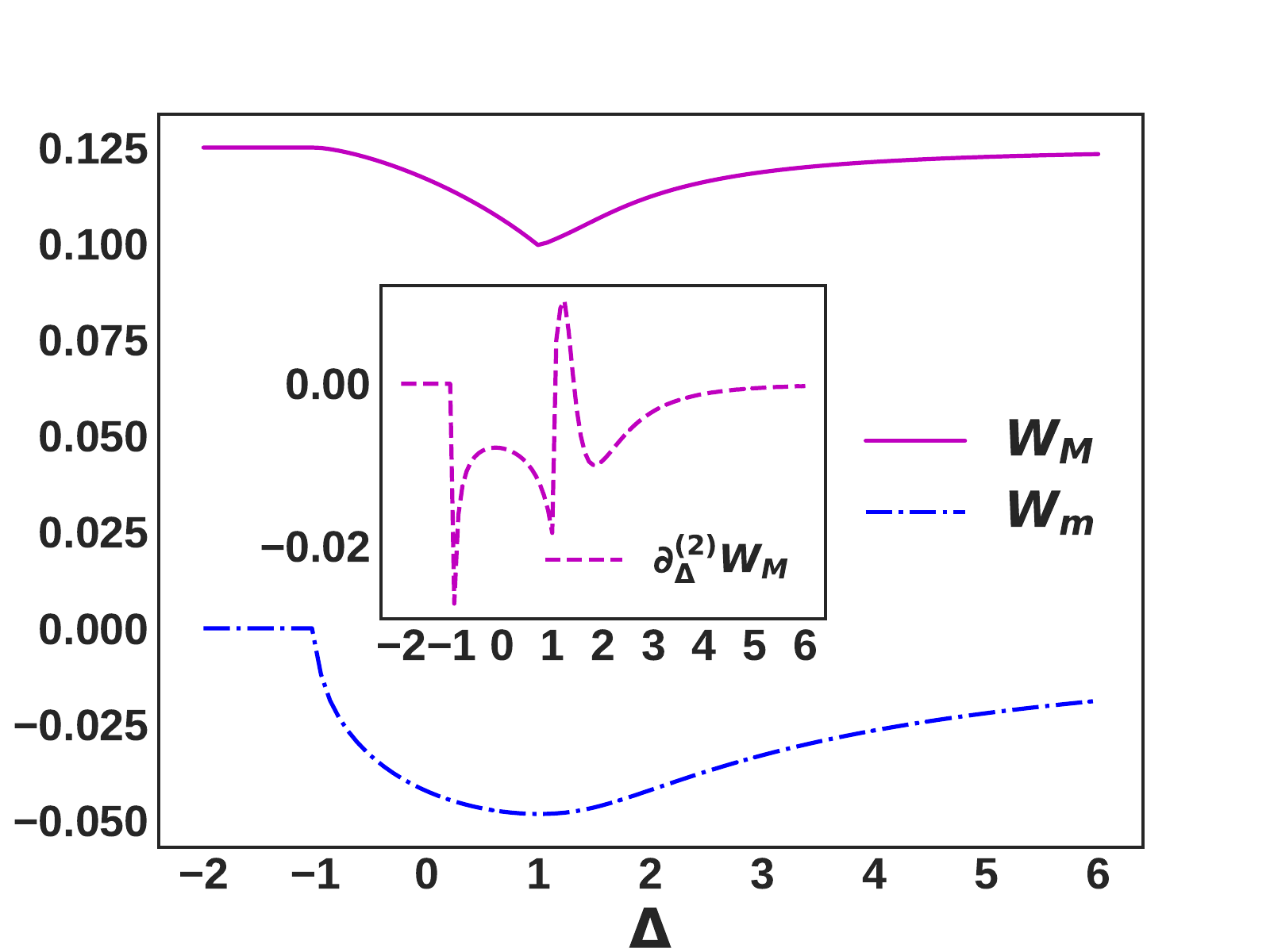}\includegraphics[width=0.5\columnwidth]{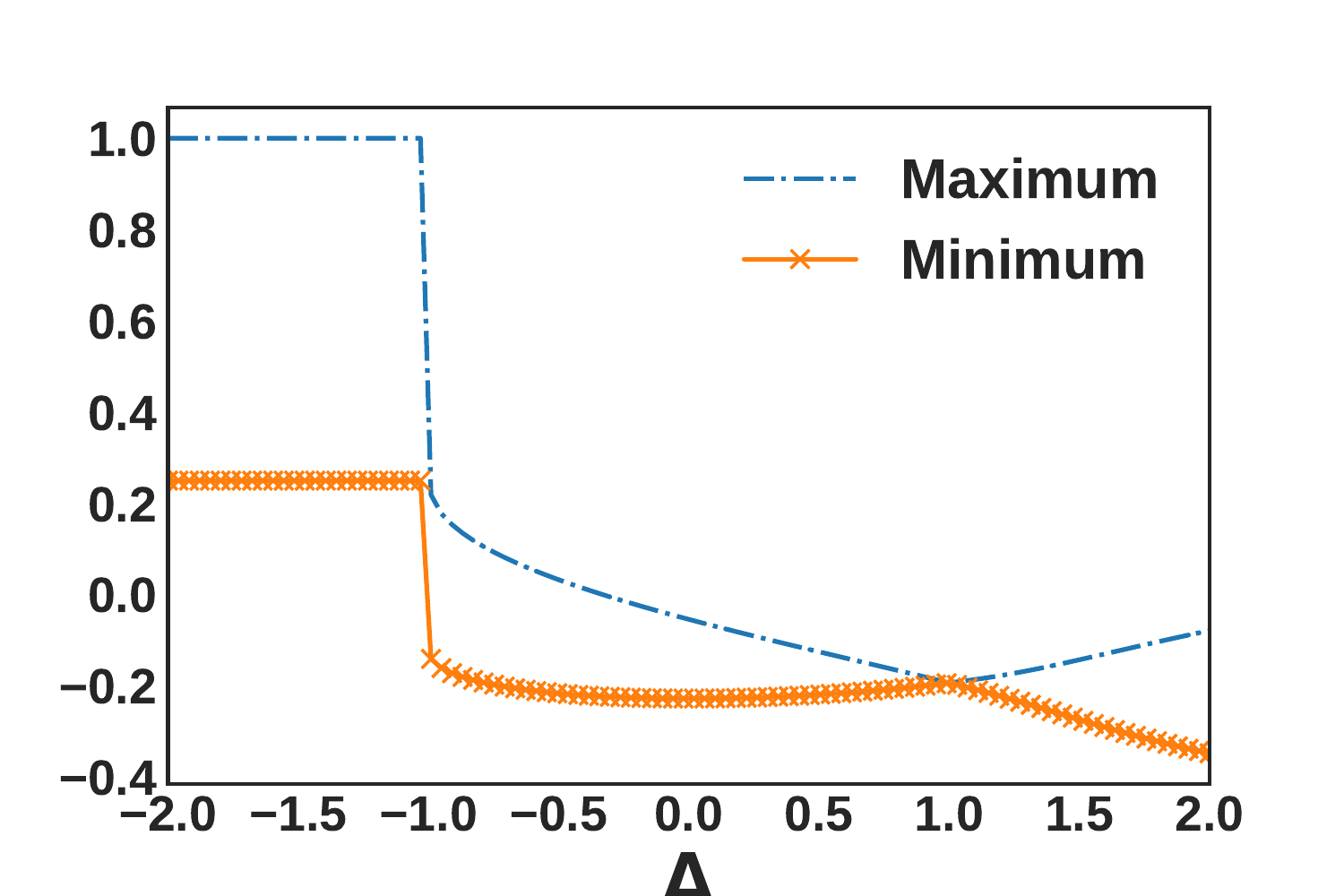}\\
\caption{(a) Extremization of the DWF following Eq.~\eqref{eqExtreme}. The topmost curve shows the maximization $W_M$ and the lower curve shows the minimization $W_m$. The inset shows the second derivative with respect to $\Delta$ of $W_M$. (b) The maximum and the minimum behavior of the GWF for the $XXZ$ model, Eq.~\eqref{gwf_xxz}.}
\label{fig3}
\end{figure}

Several correlation measures involve a minimization or maximization to be performed and often such correlation measures standout as the preferred figures of merit for studying criticality~\cite{EPL2011, SarandyPRA2009, CampbellPRA2013, Werlang2010, CakmakPRB2014}. In this regard it is interesting to consider a similar extremization procedure for the DWF. Let $W_M$ and $W_m$ be the maximized and minimized DWF over the discrete phase space, respectively, given by
\begin{equation}
\begin{aligned}
W_M&=\max(W_{00,00},W_{00,01},W_{01,00}),\\
W_m&=\min(W_{00,00},W_{00,01},W_{01,00}),
\end{aligned}
\label{eqExtreme}
\end{equation}
where we have chosen $W_{00,00}$, $W_{00,01}$, and $W_{01,00}$ to capture the three distinct behaviors exhibited in the discrete phase space. In Fig.~\ref{fig3}(a) we see $W_M$ reveals a cusp exactly at the CQPT and thus its first (second) derivative is discontinuous (divergent) at the critical point, $\Delta\!=\!1$, as shown in the inset. This indicates that the DWF could be a good alternative to correlation measures that involve extremization procedures due to the comparative simplicity in its calculation and its easy physical interpretation following Eq.~\eqref{eq4}. Looking at $\partial_\Delta^2 W_M$ of the DWF in the inset of Fig. \ref{fig3}(a) and the corresponding second derivatives of the distinct behaviors in discrete phase space shown in Fig.~\ref{fig_XXZ}, where we have destructive interference between the terms that control the DWF (for example the point (00,01)), we find that both behave quite similarly. Therefore, it appears that to be able to detect \textit{reliably} the CQPT, one requires a figure of merit that includes all the spin-spin correlation functions of the quantum system. This is further evidenced by the fact that the other parts of the phase space, where only a single spin-spin correlation term is dominant, are less sensitive to this QPT. 

We finally consider the two site GWF
\begin{align}
\text{GWF}_{\rho_{ij}}&(\theta_i,\varphi_i,\theta_j,\varphi_j)\!=\!\frac{1}{4} \Big( 1+ 3 \cos{2\theta_i}\cos{2\theta_j}  \langle \sigma^z_i\sigma^z_{j} \rangle + \nonumber\\  & 3\sin{2\theta_i}\sin{2\theta_j}\cos{2(\varphi_i-\varphi_j)} \langle \sigma^x_i\sigma^x_{j} \rangle
\Big).
\label{gwf_xxz}
\end{align}
Similarly with the $XY$ model, we are interested in a set of angles $(\theta_i,\varphi_i,\theta_j,\varphi_j)$ that yield the maximum ($\theta_i\!=\!\theta_j\!=\!\pi/2$ for $\Delta\!\leq\!1$ and $\theta_i\!=\!\theta_j\!=\!\pi/4$ for $\Delta\!>\!1$) and the minimum ($\theta_i\!=\!\theta_j\!=\!\pi/4$ for $\Delta\!\leq\!1$ and $\theta_i\!=\!\theta_j\!=\!\pi/2$ for $\Delta\!>\!1$) behavior of the GWF. One may notice that this procedure does not depend on the angles $(\varphi_i,\varphi_j)$, this is because they cancel out in Eq.~\eqref{gwf_xxz} when they are bounded by the same interval. Looking at Fig.~\ref{fig3}(b) we see that the maximum (minimum) behavior of the GWF of the $XXZ$ model is constant when $\Delta \!\!<\!\!-1$ and shows an inherited discontinuity from the spin-spin correlation functions, at the 1QPT point $\Delta\!\!=\!\!-1$. Moreover, reaching the point of the CQPT $\Delta\!=\!+1$, the angles describing the maximum (minimum) behaviors of Eq.~\eqref{gwf_xxz} switch from $\pi/4$ to $\pi/2$ ($\pi/2$ to $\pi/4$) which manifests as a cusp, revealing the CQPT at $\Delta\!\!=\!\!1$ and thus by exploiting an extremization procedure we are able to faithfully spotlight the CQPT.

\section{Conclusion \label{sec5}}
We have presented an alternative method to study quantum phase transitions from a phase space perspective using two approaches: the discrete Wigner function (DWF) and the generalized Wigner function (GWF). By establishing a connection between the phase space techniques and the thermodynamical quantities of a quantum spin-$\frac{1}{2}$ chain, we have shown the DWF and the GWF to be versatile tools in studying first, second, and infinite-order quantum phase transitions. Furthermore, we have shown that signatures of ground state factorization is only present in bipartite quantities. In addition, our approach may provide a promising tool for the experimental investigation of quantum phase transitions following the procedures proposed in Refs.~\cite{expwf, expdwf2, GWF_entanglement_PRA}. Furthermore, through~\Cref{dwf_single_site,eq4,gwf_ss,gwf_ij,gwf_ijk}, a given DWF/GWF is easily physically interpreted and can be generalized to higher dimensional systems which is a task proven to be difficult and complex for quantum correlations measures. Beyond characterizing phase transitions, our approach also provides insight into the behavior of various correlation measures and quantum coherence in such systems. While we have focused on equilibrium systems, we expect our approach to be useful in examining the dynamical properties of such critical systems~\cite{CampbellPRB2016, HeylReview, NJPSchachenmayer, PRASchachenmayer, PRBGasenzer, QST2019, PRXSchachenmayer}. 

\section*{Acknowledgements}
ZM and MEB would like to thank Fabio Benatti, Marcello Dalmonte, Rosario Fazio, and Ugo Marzolino for valuable comments and discussions. SC is grateful to Tony Apollaro and Bar\i\c{s} \c{C}akmak for fruitful discussions. SC gratefully acknowledges the Science Foundation Ireland Starting Investigator Research Grant ``SpeedDemon" (No. 18/SIRG/5508) for financial support. The calculation were done using NumPy~\cite{numpy}, SciPy~\cite{scipy}, Matplotlib~\cite{matplotlib} and QuTiP~\cite{qutip1, qutip2}

\bibliography{Manuscript}

\section*{Appendix}
Here we report the explicit form of the GWF for a three sites of the $XY$ chain where we have used Wick's theorem~\cite{Wick} to evaluate the three-spin correlation functions finding
\onecolumngrid
	\begin{align}
	&\text{GWF}_{\rho_{ijk}}(\theta_i,\varphi_i,\theta_j,\varphi_j,\theta_k,\varphi_k) = \frac{1}{8} \Big[ 1-\sqrt{3}\left(\cos{2\theta_i} + \cos{2\theta_j} + \cos{2\theta_k} \right) \langle \sigma^z \rangle +
	3 \cos{2\varphi_i}\sin{2\theta_i}\cos{2\varphi_k}\sin{2\theta_k} \langle \sigma^x_i\sigma^x_k \rangle  \nonumber \\& 3\left( \cos{2\varphi_i}\sin{2\theta_i}\cos{2\varphi_j}\sin{2\theta_j} + \cos{2\varphi_j}\sin{2\theta_j}\cos{2\varphi_k}\sin{2\theta_k}  \right) \langle \sigma^x_i\sigma^x_j \rangle +
	3 \sin{2\theta_i}\sin{2\varphi_i} \sin{2\theta_k}\sin{2\varphi_k} \langle \sigma^y_i\sigma^y_{k} \rangle + \nonumber \\& + 3 \left(\sin{2\theta_i}\sin{2\theta_j}\sin{2\varphi_i}\sin{2\varphi_j} + \sin{2\theta_i}\sin{2\theta_k}\sin{2\varphi_j}\sin{2\varphi_k} \right) \langle \sigma^y_i\sigma^y_{k} \rangle +
	3 \cos{2\theta_i}\cos{2\theta_k}  \langle \sigma^z_i\sigma^z_{k} \rangle \nonumber \\& +3 \left(\cos{2\theta_i}\cos{2\theta_j} + \cos{2\theta_j}\cos{2\theta_k} \right) \langle \sigma^z_i\sigma^z_{j} \rangle -
	3\sqrt{3} \cos{2\varphi_i}\sin{2\theta_i}\cos{2\varphi_j}\sin{2\theta_j}\cos{2\theta_k}   \langle \sigma^x_i\sigma^x_{j} \rangle  \langle \sigma^z \rangle + \nonumber \\ &
	3\sqrt{3} \cos{2\varphi_i}\sin{2\theta_i}\cos{2\varphi_k}\sin{2\theta_k}\cos{2\theta_j}   \langle \sigma^x_i\sigma^x_{k} \rangle  \langle \sigma^z \rangle - 3\sqrt{3} \sin{2\theta_i}\sin{2\varphi_i} \sin{2\theta_j}\sin{2\varphi_j}\cos{2\theta_k}   \langle \sigma^y_i\sigma^y_{j} \rangle  \langle \sigma^z \rangle + \nonumber \\&
	3\sqrt{3} \sin{2\theta_i}\sin{2\varphi_i} \sin{2\theta_k}\sin{2\varphi_k}\cos{2\theta_j}    \langle \sigma^y_i\sigma^y_{k} \rangle  \langle \sigma^z \rangle -
	3\sqrt{3} \cos{2\theta_i}\cos{2\theta_j}\cos{2\theta_k}  \left( \langle \sigma^z_i\sigma^z_{j} \rangle - \langle \sigma^z_i\sigma^z_{k} \rangle \right)  \langle \sigma^z \rangle
	\Big].
	\label{gwf_ijk}
	\end{align}
\end{document}